\documentclass[journal,twoside,web]{ieeecolor}

\usepackage[utf8]{inputenc}
\usepackage{amsmath}
\usepackage{amssymb}

\usepackage{amsthm}
\usepackage{graphicx}
\usepackage{bm}
\usepackage{dsfont}
\usepackage{subcaption}
\usepackage{tikz}

\usepackage[sort,compress]{cite}
\usepackage{amsfonts}
\usepackage{url}
\usepackage[pdftex, pdfstartview={FitV}, pdfpagelayout={TwoColumnLeft},bookmarksopen=true,plainpages = false, colorlinks=true, linkcolor=black, citecolor = black, urlcolor = blue,filecolor=black , pagebackref=false,hypertexnames=false, plainpages=false, pdfpagelabels ]{hyperref}
\usepackage{balance}
\usepackage{array}
\usepackage{tabularx}
\usepackage{booktabs}
\newcolumntype{Y}{>{\centering\arraybackslash}X}

\usepackage{tikz}
\usepackage{caption}

\usepackage[capitalize]{cleveref}  

\newtheorem{theorem}{Theorem}[section]
\newtheorem{corollary}[theorem]{Corollary}
\newtheorem{lemma}[theorem]{Lemma}

\newtheorem{definition}[theorem]{Definition}

\crefname{section}{Section}{Sections}
\crefname{theorem}{Theorem}{Theorems}
\crefname{lemma}{Lemma}{Lemmas}
\crefname{table}{Table}{Tables}
\crefformat{equation}{(#2#1#3)}
\crefname{algocf}{Algorithm}{Algorithms}
\Crefname{algocf}{Algorithm}{Algorithms}
\crefname{ALC@unique}{Line}{Lines}

\newcolumntype{M}[1]{>{\centering\arraybackslash}m{#1}}

\usepackage{accents}
\newcommand{\ubar}[1]{\underaccent{\bar}{#1}}

\usepackage[]{changes}
\definechangesauthor[name={MFE}, color={blue}]{MFE}

\definechangesauthor[name={RB}, color={green}]{RB}

\definechangesauthor[name={SO}, color={red}]{SO}

\usepackage{algorithm,algorithmic}

\newcommand{\CROWNA}{\mathbf{M}}
\newcommand{\CROWNb}{\mathbf{n}}

\usepackage{tikz}
\usetikzlibrary{shapes.misc}
\tikzset{cross/.style={cross out, draw=black, inner sep=0pt, outer sep=0pt, minimum size=4, very thick}}
\newcommand{\zb}{\mathbf{z}}
\newcommand{\xb}{\mathbf{x}}
\newcommand{\nb}{\mathbf{n}}
\newcommand{\lb}{\mathbf{l}}
\newcommand{\ub}{\mathbf{u}}
\newcommand{\vb}{\mathbf{v}}

\newcommand{\alphab}{\bm{\alpha}}
\newcommand{\betab}{\bm{\beta}}

\newcommand{\istate}{k}
\usepackage{bbold}

\usepackage{mathtools}
\DeclarePairedDelimiter\floor{\lfloor}{\rfloor}

\usepackage{lcsys}
\usepackage{textcomp}
\def\BibTeX{{\rm B\kern-.05em{\sc i\kern-.025em b}\kern-.08em
    T\kern-.1667em\lower.7ex\hbox{E}\kern-.125emX}}
\begin{document}
\title{DRIP: Domain Refinement Iteration with Polytopes for Backward Reachability Analysis of Neural Feedback Loops}
\author{Michael Everett, \IEEEmembership{Member, IEEE}, Rudy Bunel, Shayegan Omidshafiei
\thanks{This manuscript was first submitted for review on 12/8/2022.}
\thanks{M. Everett conducted this research while at Google Research and is now with Northeastern University, Boston, MA 02118 USA (e-mail: m.everett@northeastern.edu). }
\thanks{R. Bunel is with DeepMind, London, EC4A 3TW UK (e-mail: rbunel@google.com).}
\thanks{S. Omidshafiei is with 
Google Research, Cambridge, MA 02139 USA (e-mail: somidshafiei@google.com).}}

\maketitle

\begin{abstract}
Safety certification of data-driven control techniques remains a major open problem.
This work investigates backward reachability as a framework for providing collision avoidance guarantees for systems controlled by neural network (NN) policies.
Because NNs are typically not invertible, existing methods conservatively assume a domain over which to relax the NN, which causes loose over-approximations of the set of states that could lead the system into the obstacle (i.e., backprojection (BP) sets).
To address this issue, we introduce DRIP, an algorithm with a refinement loop on the relaxation domain, which substantially tightens the BP set bounds.
Furthermore, we introduce a formulation that enables directly obtaining closed-form representations of polytopes to bound the BP sets tighter than prior work, which required solving linear programs and using hyper-rectangles.
Furthermore, this work extends the NN relaxation algorithm to handle polytope domains, which further tightens the bounds on BP sets.
DRIP is demonstrated in numerical experiments on control systems, including a ground robot controlled by a learned NN obstacle avoidance policy.
\end{abstract}


\section{Introduction}
\IEEEPARstart{N}{eural} networks (NNs) offer promising capabilities for data-driven control of complex systems (e.g., self-driving cars).
However, formally certifying safety properties of systems that are controlled by NNs remains an open challenge.

To this end, recent work developed reachability analysis techniques for Neural Feedback Loops (NFLs), i.e., dynamical systems with NN control policies~\cite{everett2021reachability,hu2020reach,chen2022one,sidrane2021overt}.
For \textit{forward} reachability analysis, these techniques compute the set of states the system could reach in the future, given an initial state set, trained NN, and dynamics model.
Due to the NNs' high dimensionality and nonlinearities, exact analysis is often intractable. 
Thus, current methods instead compute guaranteed over-approximations of the reachable sets, based on relaxations of the nonlinearities in the NN~\cite{everett2021reachability,hu2020reach,chen2022one} or dynamics~\cite{sidrane2021overt,everett2021reachability}.
While forward reachable sets can then be used to check whether the closed-loop system satisfies desired properties (e.g., reaching the goal), forward methods can be overly conservative for obstacle avoidance~\cite{Rober22_CDC}.

Thus, this work focuses on \textit{backward} reachability analysis~\cite{everett2021reachability,Rober22_CDC,Rober23_ACC,Rober22_OJCSYS,vincent2021reachable,bak2022closed}.
The goal of backward reachability analysis is to compute backprojection (BP) sets, i.e., the set of states the will lead to a given target/obstacle set under the given NN control policy and system dynamics.
The system can be certified as safe if it starts outside the BP sets, but the non-invertibility of NNs presents a major challenge in calculating these sets.
Recent work~\cite{everett2021reachability,Rober22_CDC,Rober23_ACC,Rober22_OJCSYS} proposes to first compute a backreachable (BR) set, i.e., the set of states that lead to the target set for some control input within known control limits.
These methods then relax the NN controller over the BR set, which is a superset of the BP set, and calculate bounds on the BP set via linear programs.


However, a key challenge with that formulation is the BP set over-approximations can remain loose.
A major cause of this conservativeness is that the BR set is often large, leading to loose NN relaxations, and thus loose BP set approximations.
Moreover, conservativeness compounds when calculated over multiple timesteps due to the wrapping effect~\cite{neumaier1993wrapping}.
Prior work introduced strategies based on set partitioning~\cite{Rober22_CDC,Rober22_OJCSYS,Rober23_ACC} and mixed integer linear programming (MILP)~\cite{Rober22_OJCSYS} to improve tightness, but these methods are fundamentally hindered by initializing with the BR set.
This paper instead proposes a refinement loop, which, for a particular timestep, iteratively uses the previous BP set estimate to relax the NFL and compute a new BP set, which can lead to much tighter BP set estimates.
At each iteration, this new strategy shrinks the domain over which the NFL is relaxed, which leads to a less conservative relaxation and ultimately tighter bounds on the BP sets.
If desired, this idea could be used in conjunction with set partitioning~\cite{Rober22_CDC,Rober22_OJCSYS,Rober23_ACC}.

Another key limitation of prior work is the use of axis-aligned bounding boxes to represent BP set estimates~\cite{Rober22_CDC,Rober22_OJCSYS,Rober23_ACC}.
These hyperrectangles are computed by solving several linear programs (LPs) over states and controls, but hyperrectangles are often a poor approximation of the true BP sets.
Instead, this paper introduces a new approach that enables directly obtaining a closed-form polytope representation for the BP set estimate.
A key reason why this enables tighter BP set bounds is that the facets of the target set (or future BP sets in the multi-timestep algorithm) can be directly used as the objective matrix for the relaxation algorithm, which also automatically addresses the common issue of choosing facet directions in polytope design.
Furthermore, we show that the NN can be relaxed over these polytope BP sets, which leads to much tighter relaxations and BP set estimates compared to prior work, which inflated polytopes to hyperrectangles.

To summarize, the main contribution of this work is DRIP, an algorithm that provides safety guarantees for NFLs, which includes domain refinement and a polytope formulation to give tighter certificates than state-of-the-art methods.
Numerical experiments on the same data-driven control systems as in prior state-of-the-art works~\cite{Rober22_CDC,Rober22_OJCSYS,Rober23_ACC} demonstrate that DRIP can provide $371\times$ tighter bounds while remaining computationally efficient (${\sim0.3}$s) and can certify obstacle avoidance for a robot with a learned NN policy.

\section{Background}
This section defines the dynamics, relaxations, and sets used for backward reachability analysis.

\subsection{NFL Dynamics}
As in~\cite{Rober22_OJCSYS}, we assume linear time-invariant (LTI) dynamics,
\begin{align}
\begin{split}
    \mathbf{x}_{t+1} & = \mathbf{A} \mathbf{x}_{t} + \mathbf{B} \mathbf{u}_t + \mathbf{c} \triangleq f(\mathbf{x}_{t}, \mathbf{u}_t) \label{eqn:lti_dynamics} \, ,
\end{split}
\end{align}
where $\mathbf{x}_{t} \in \mathds{R}^{n_x}$ is the state at discrete timestep $t$, $\mathbf{u}_t \in \mathds{R}^{n_u}$ is the input, $\mathbf{A} \in \mathds{R}^{n_x \times n_x}$ and $\mathbf{B} \in \mathds{R}^{n_x \times n_u}$ are known system matrices, and $\mathbf{c} \in \mathds{R}^{n_x}$ is a known exogenous input. 
We assume $\mathbf{x}_t \in \mathcal{X}\subseteq\mathds{R}^{n_x}$ and $\mathbf{u}_t \in \mathcal{U}\subseteq\mathds{R}^{n_u}$, where $\mathcal{X}$ and $\mathcal{U}$ are convex sets defining the operating region of the state space and control limits, respectively.
The closed-loop dynamics are 
\begin{align}
\begin{split}
    \mathbf{x}_{t+1} & = \mathbf{A} \mathbf{x}_{t} + \mathbf{B}\pi(\mathbf{x}_t) + \mathbf{c} \triangleq p(\mathbf{x}_t;\pi) \, ,
    \label{eqn:nfl}
\end{split}
\end{align}
where $\pi(\cdot)$ is a state-feedback control policy, discussed next.

\subsection{NN Control Policy \& Relaxation}\label{sec:background:nn_policy_and_relaxation}

The control policy, $\pi: \mathcal{X} \to \mathcal{U}$, can be an arbitrary computation graph,
composed of primitives that can be linearly relaxed~\cite{xu2020automatic,dathathri2020enabling,jaxverify}.
For example, given an $m$-layer NN with $n_i$ neurons in layer $i$ and nonlinear activation $\sigma_i: \mathds{R}^{n} \to \mathds{R}^{n}$ such that post-activation $\zb_i = \sigma_i(\zb_{i-1})$ (e.g., ReLU, sigmoid), we need to be able to define $\alphab_{l,i}, \betab_{l,i}, \alphab_{u,i}, \betab_{u,i}$ such that 
\begin{equation}
    \alphab_{l, i} \zb_{i-1} + \betab_{l,i} \leq \zb_i \leq \alphab_{u, i} \zb_{i-1} + \betab_{u,i} \, .
    \label{eq:lin_relaxation_zi}
\end{equation}
For each primitive, values of $\alphab_{l,i}, \betab_{l,i}, \alphab_{u,i}, \betab_{u,i}$ depend on the algorithm employed and are derived from intermediate bounds over the activation of the network, which can be obtained by applying the bound computation procedure (described next) to subsets of the network.

CROWN~\cite{zhang2018efficient} and LIRPA algorithms in general~\cite{xu2020automatic} enable backward propagation of bounds (from function output to input, not to be confused with backward in time).
For example, consider evaluating a lower bound on $\mathbf{c}\cdot\zb_i + b $, using the relaxations defined in~\cref{eq:lin_relaxation_zi} to replace $\zb_i$:
  \begin{align}
    &\min \mathbf{c} \cdot \zb_i + b \\ \nonumber
    \geq & \min\  [\mathbf{c}]^+ \left(\alphab_{l, i} \zb_{i-1} + \betab_{l,i}\right) + [\mathbf{c}]^- \left(\alphab_{u, i} \zb_{i-1} + \betab_{u,i}\right) + b\\ \nonumber
    = & \min \left([\mathbf{c}]^+\alphab_{l, i}  + [\mathbf{c}]^- \alphab_{u, i}\right) \zb_{i-1} + [\mathbf{c}]^+\betab_{l, i}  + [\mathbf{c}]^- \betab_{u, i} + b,
  \end{align}
where \mbox{$[\mathbf{c}]^+ = \max(\mathbf{c}, 0)$} and \mbox{$[\mathbf{c}]^- = \min(\mathbf{c}, 0)$}.
Through backsubstitution, we transformed a lower bound defined as a linear function over $\zb_i$ into a lower bound defined as a linear function over $\zb_{i-1}$.
By repeated application of this procedure to all operations in the network, we can obtain bounds that are affine in the input.
That is, for a computation graph $p$ from~\cref{eqn:nfl}, with input $\mathbf{x}_t$ and output $\mathbf{x}_{t+1}$, given an objective matrix $\mathbf{C}\in \mathds{R}^{n_\text{facets} \times n_x}$, we compute $\mathbf{M}\in\mathds{R}^{n_\text{facets} \times n_x}$ and $\mathbf{n}\in\mathds{R}^{n_\text{facets}}$ such that
\begin{equation}
     \mathbf{M} \mathbf{x}_{t} + \mathbf{n} \leq \mathbf{C} \mathbf{x}_{t+1}. \label{eqn:affine_bounds_crown}
\end{equation}
If $\mathbf{x}_t$ is within some known set $\mathcal{X}_t$, these bounds can be concretized by solving \mbox{$\min_{\mathbf{x}_t \in \mathcal{X}_t} \mathbf{M} \mathbf{x}_t + \mathbf{n}$}.
Closed-form solutions exist for some forms of $\mathcal{X}_t$.

Prior CROWN-based backward reachability works~\cite{Rober22_CDC,Rober22_OJCSYS,Rober23_ACC} relaxed the control policy $\pi$, whereas this work directly relaxes the closed-loop dynamics, $p$, drawing inspiration from~\cite{chen2022one}, which focused on forward reachability.

\subsection{Backreachable Sets \& Backprojection Sets}
For target set $\mathcal{X}_T \subseteq \mathcal{X}$, define $t$-step BP and BR sets as
\begin{align}
    \mathcal{P}_{-t}(\mathcal{X}_T; \pi) &\triangleq \{ \mathbf{x} \lvert\ p^{t}(\mathbf{x}; \pi) \in \mathcal{X}_T \}
    \label{eqn:backprojection_sets} \\
    \mathcal{R}_{t}(\mathcal{X}_T) &\triangleq \{ \mathbf{x}\ \lvert\ \exists \pi': \mathcal{X} \to \mathcal{U} \mathrm{\ s.t.\ }
     p^{t}(\mathbf{x}; \pi') \in \mathcal{X}_T \}. \nonumber
\end{align}
where $t>0$ and $p^{t+1} \triangleq p \circ p^{t}$.
We will typically drop the arguments and simply write $\mathcal{R}_{t}, \mathcal{P}_{t}$.
Clearly, $\mathcal{P}_{t} \subseteq \mathcal{R}_{t}$.
Since exactly computing BP sets is computationally difficult, this work aims to compute BP over-approximations (BPOAs), which provide a (guaranteed) conservative description of all states that will lead to the target set.
For sets, upper bar notation conveys an outer bound, e.g., $\bar{\mathcal{A}} \supseteq \mathcal{A}$; for vectors, bar notation conveys upper/lower bounds, e.g., $\ubar{\mathbf{a}} \leq \mathbf{a} \leq \bar{\mathbf{a}}$.

\begin{definition}
For some $t<0$, $\bar{\mathcal{P}}_{t}$ is a BPOA if $\bar{\mathcal{P}}_{t} \supseteq \mathcal{P}_{t}$.
\end{definition}




One efficient way to outer bound $\mathcal{R}_{-1}$ with hyper-rectangle ${[\ubar{\mathbf{x}}_{-1}, \bar{\mathbf{x}}_{-1}]}$ is to solve the following optimization problems for each state $\istate \in [n_x]$, with the $i$-th standard basis vector, $\mathbf{e}_{i}$~\cite{everett2021reachability}:
\begin{equation}
\begin{split}
& \bar{\mathbf{x}}_{-1; \istate} = \max_{\mathbf{x} \in \mathcal{X},\mathbf{u}\in\mathcal{U}\ \text{s.t.}\ f(\mathbf{x}, \mathbf{u}) \in \mathcal{X}_T} \mathbf{e}_k^\top \mathbf{x} \\ & \ubar{\mathbf{x}}_{-1; \istate} = \min_{\mathbf{x} \in \mathcal{X},\mathbf{u}\in\mathcal{U}\ \text{s.t.}\ f(\mathbf{x}, \mathbf{u}) \in \mathcal{X}_T} \mathbf{e}_k^\top \mathbf{x}. \label{eqn:lp_obj}
\end{split}
\end{equation}

\section{Approach}
This section introduces our proposed approach, Domain Refinement Iteration with Polytopes (DRIP), with descriptions of the 3 technical contributions in \cref{sec:approach:refinement_loop,sec:approach:polytope_formulation,sec:approach:polytope_input_bounds} and a summary of the algorithm in \cref{sec:approach:algorithm}.

\subsection{Improved BP Set Representation: Polytope Formulation}\label{sec:approach:polytope_formulation}

\begin{figure*}[t]
    \centering
    \includegraphics[page=5, width=0.8\linewidth,clip, trim=0 150 0 0]{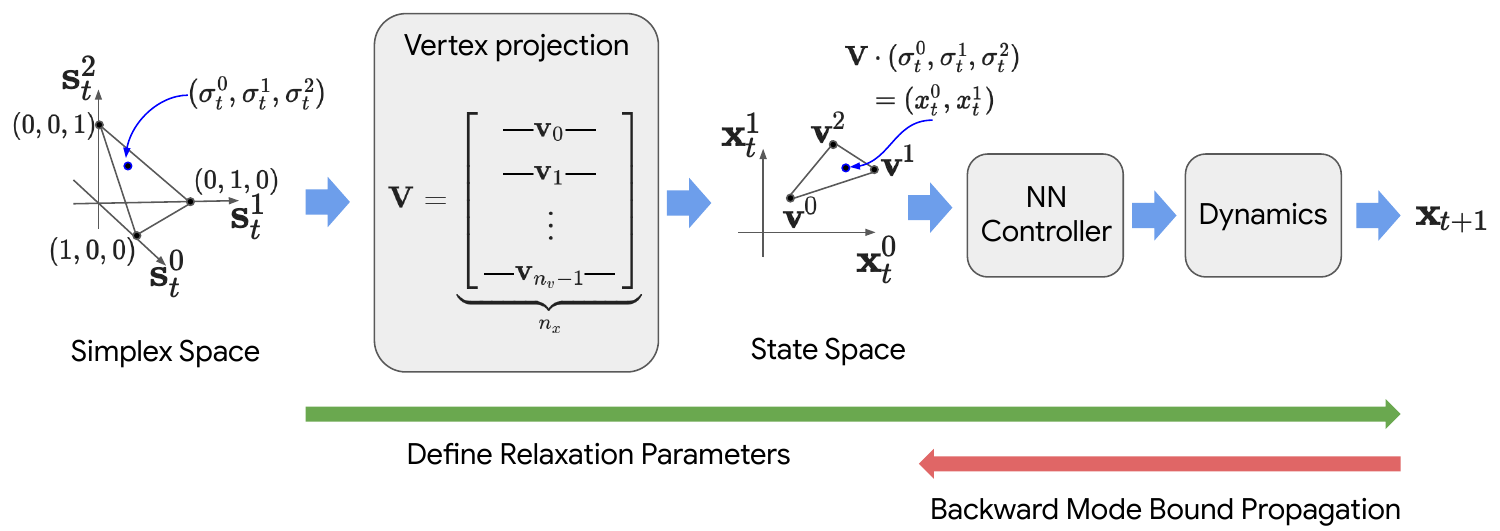}
    \caption{Architecture to enable relaxation of closed-loop dynamics over a polytope description of the current state set. A linear transformation, $\mathbf{V}$ (stacked polytope vertices, $n_v=3$ here), is added to the front of the closed-loop dynamics. This transformation projects the standard $n_v$-simplex onto the state space, leading to a polytope of states over which the closed-loop dynamics is relaxed. During the relaxation, the function including the vertex projection is used to define relaxation parameters, but the backward mode bound propagation stops before the vertex projection to compute $\CROWNA \mathbf{x}_t + \CROWNb \leq \mathbf{C}\mathbf{x}_{t+1}$.}
    \label{fig:simplex_polytope}
    \vspace{-0.2in}
\end{figure*}

Whereas prior work (e.g., \cite{Rober22_OJCSYS}) represents BPOAs as hyper-rectangles, this work introduces a formulation based on halfspace-representation (H-Rep) polytopes.

\begin{lemma}\label{thm:bpoa_polytope}
Given closed-loop dynamics $p$~\cref{eqn:nfl}, $\mathbf{A}_T, \mathbf{b}_T$ parameterizing target set $\mathcal{X}_T = \{ \mathbf{x} \ \lvert \mathbf{A}_T\mathbf{x} \leq \mathbf{b}_T \}$, and $\mathbf{A}_\text{relax}, \mathbf{b}_\text{relax}$ parameterizing a BPOA, $\bar{\mathcal{P}}_{-1}' = \{ \mathbf{x} \ \lvert \mathbf{A}_\text{relax}\mathbf{x} \leq \mathbf{b}_\text{relax} \} \supseteq \mathcal{P}_{-1}$, the following set is also a BPOA,
\begin{align}
    \bar{\mathcal{P}}_{-1} &= \left\{ \mathbf{x}\ \lvert \begin{bmatrix} \mathbf{M} \\ \mathbf{A}_\text{relax} \end{bmatrix} \mathbf{x} \leq \begin{bmatrix} \mathbf{b}_T - \mathbf{n} \\ \mathbf{b}_\text{relax} \end{bmatrix} \right\}, \label{eqn:bpoa_as_polytope}
\end{align}
with $\bar{\mathcal{P}}'_{-1} \supseteq \bar{\mathcal{P}}_{-1} \supseteq \mathcal{P}_{-1}$, and with $\mathbf{M}, \mathbf{n}$ defined below.

\begin{proof}
Run CROWN~\cref{eqn:affine_bounds_crown} on $p$ over the domain $\bar{\mathcal{P}}_{-1}'$ with objective $\mathbf{C}=\mathbf{A}_T$ to obtain $\CROWNA, \CROWNb$ such that
\begin{align}
    \CROWNA \mathbf{x}_{t-1} + \CROWNb &\leq \mathbf{A}_T \mathbf{x}_{t} \leq \mathbf{b_T} \quad \forall \mathbf{x}_{t-1} \in \bar{\mathcal{P}}'_{-1} \label{eqn:bpoa_inequality}.
\end{align}
Thus, the set of $\mathbf{x}_{t-1}$ that lead to $\mathcal{X}_T$ in one timestep is bounded by the constraints \cref{eqn:bpoa_inequality} and $\mathbf{x}_{t-1} \in \bar{\mathcal{P}}'_{-1}$.
$\bar{\mathcal{P}}'_{-1} \supseteq \bar{\mathcal{P}}_{-1}$ by construction, and $\bar{\mathcal{P}}_{-1} \supseteq \mathcal{P}_{-1}$ since $\mathcal{P}_{-1}$ could contain $\mathbf{x}_{t-1}$ that do not satisfy  \cref{eqn:bpoa_inequality} due to the relaxation gap. 
\end{proof}
\end{lemma}

\begin{corollary}\label{thm:bpoa_polytope_br}
To find a BPOA without being provided $\mathbf{A}_\text{relax}, \mathbf{b}_\text{relax}$, solve~\cref{eqn:lp_obj} to get hyper-rectangular bounds, $\bar{\mathcal{R}}_{-1} \supseteq \mathcal{R}_{-1}$. $\bar{\mathcal{R}}_{-1}$ is also a BPOA and can be used for~\cref{thm:bpoa_polytope}.
\end{corollary}

\subsection{Improved Relaxation Domain: Refinement Loop}\label{sec:approach:refinement_loop}

While $\bar{\mathcal{R}}_{-1}$ provides a domain for relaxing the NN, this domain is often excessively conservative and leads to large BPOAs.
To address this issue, this work leverages the following to tighten the domain used for relaxation.
\begin{corollary}\label{thm:dri}
Given closed-loop dynamics $p$ from~\cref{eqn:nfl}, $\mathbf{A}_T, \mathbf{b}_T$ that parameterize $\mathcal{X}_T = \{ \mathbf{x} \ \lvert \mathbf{A}_T\mathbf{x} \leq \mathbf{b}_T \}$, and initial BPOA $\bar{\mathcal{P}}^{0}_{-1}$ (e.g., using \cref{thm:bpoa_polytope_br}), applying \cref{thm:bpoa_polytope} iteratively for $n_\text{iters}$ results in the sequence of BPOAs, $\bar{\mathcal{P}}^{0}_{-1} \supseteq \bar{\mathcal{P}}^{1}_{-1} \supseteq \cdots \supseteq \bar{\mathcal{P}}^{n_\text{iters}}_{-1} \supseteq \mathcal{P}_{-1}$.
\end{corollary}

This iterative refinement can be performed, for example, for a specified number of steps or until the reduction in BPOA volume between refinement steps reaches some threshold.


\setlength{\textfloatsep}{4pt}
\begin{algorithm}[t]
 \caption{CROWNSimplex}
 \begin{algorithmic}[1]
 \setcounter{ALC@unique}{0}
 \renewcommand{\algorithmicrequire}{\textbf{Input:}}
 \renewcommand{\algorithmicensure}{\textbf{Output:}}
 \REQUIRE closed-loop dynamics $f$, polytope vertices $\mathbf{V}$, objective matrix $\mathbf{C}$
 \ENSURE $\CROWNA$, $\CROWNb$, describing affine bounds from $\mathbf{x}_t$ to $\mathbf{x}_{t+1}$
 \STATE $\mathbf{s} \in \Delta_{n_v}$ \texttt{\# input to computation graph}
 \STATE $\mathbf{x}_{t} = \mathbf{V} \cdot \mathbf{s}$
 \STATE $\mathbf{x}_{t+1} = f_{\text{simplex}}(\mathbf{s}) = f(\mathbf{V}\cdot \mathbf{s})$ \label{alg:crown_simplex:add_layer}
 \STATE $\alphab, \betab \leftarrow \text{defineRelaxationParams}(f_\text{simplex}, [\mathbb{0}_{n_v}, \mathbb{1}_{n_v}], 1)$\label{alg:crown_simplex:fwd_bounds}
 \STATE $\CROWNA, \CROWNb \leftarrow \text{bwdModeProp}(f_\text{simplex}, \mathbf{x}_{t+1} \rightarrow \mathbf{x}_t, \alphab, \betab, \mathbf{C})$\label{alg:crown_simplex:bwd_bounds}
 \RETURN $\CROWNA, \CROWNb$ 
 \end{algorithmic}\label{alg:crown_simplex}
\end{algorithm}

\subsection{Improved Relaxation over BP Set: Polytope Input Bounds}\label{sec:approach:polytope_input_bounds}

Existing algorithms~\cite{everett2021reachability,Rober22_CDC,Rober22_OJCSYS,Rober23_ACC} simplified the bound computation by assuming the domain over which to relax the network was a hyper-rectangle \mbox{$\mathcal{X} = [\mathbf{l}, \mathbf{u}]^d$}, which enables
\begin{equation}
    \min_{x \in \mathcal{X}}\ \mathbf{M} \xb + \nb = [\mathbf{M}]^+ \lb + [\mathbf{M}]^- \ub + \nb \, .  
\end{equation}
Assuming that the input domain is a convex polytope $\mathcal{C}$, the strategy consists in first solving $2d$ LPs, $l_i = \min_{x \in \mathcal{C}} x_i$ and $u_i = \max_{x \in \mathcal{C}} x_i$, to obtain the hyperrectangular bounds on $\mathcal{C}$, which may introduce conservativeness.

Instead, the simplex is another domain that enables efficient concretization of linear bounds, but also enables natural representation of a convex polytope.
For $\mathcal{X} = \Delta_d = \left\{\xb \in \mathcal{R}^d  |\  \xb \geq 0, \sum_i^d x_i = 1\right\}$,
\begin{equation}
    \min_{x \in \mathcal{X}}\ \mathbf{M} \xb + \nb = \mathbf{M}_{\text{argmin[M]}} + \nb \,,
\end{equation}
where $\mathbf{M}_{\text{argmin[M]}}$ is a vector of the minimum of each row of $\mathbf{M}$.
We can represent a convex polytope $\mathcal{C}$, with a simplex and a transformation based on its vertices $\left\{\vb_0, \vb_1, \dotsc, \vb_n\right\}$:
\begin{equation}
    \xb \in \mathcal{C} \iff \ \exists \mathbf{s} \in \Delta_d\ \text{such that } \xb = \mathbf{V} \cdot \mathbf{s} \, ,
\end{equation}
where $\mathbf{V} = [\vb_0^T; \vb_1^T; \dotsc; \vb_{n-1}^T] \in \mathds{R}^{n_v \times n_x}$. 
\cref{fig:simplex_polytope} shows how this is integrated in the bound propagation: the multiplication by the matrix of vertices is simply prepended to the NN controller as an initial linear layer (an exact representation of polytope $\mathcal{C}$).

\setlength{\textfloatsep}{2pt}
\begin{algorithm}[t]
 \caption{Domain Refinement with Polytopes (DRIP)}
 \begin{algorithmic}[1]
 \setcounter{ALC@unique}{0}
 \renewcommand{\algorithmicrequire}{\textbf{Input:}}
 \renewcommand{\algorithmicensure}{\textbf{Output:}}
 \REQUIRE target set $\mathcal{X}_T$, policy $\pi$, dynamics $p$, iterations $n_\text{iters}$, algorithm variant $alg \in \{\text{DRIP, DRIP-HPoly}\}$
 \ENSURE BP set approximation $\bar{\mathcal{P}}_{-1}(\mathcal{X}_T)$
    \STATE $\mathbf{A}_{\mathcal{X}_T}, \mathbf{b}_{\mathcal{X}_T} \leftarrow \mathcal{X}_T = \{\mathbf{x}\ |\ \mathbf{A}_{\mathcal{X}_T} \mathbf{x} \leq \mathbf{b}_{\mathcal{X}_T}\}$
    \STATE $\bar{\mathcal{P}}_{-1} \leftarrow\! \bar{\mathcal{R}}_{-1} = \mathrm{backreach}(\mathcal{X}_T, \mathcal{U}, f)$ \label{alg:one_step_backprojection:backreach}
    \FOR{$i$ in $\{1, 2, \ldots, n_\text{iters}\}$} \label{alg:one_step_backprojection:loop}
        \IF{$alg == \text{DRIP}$}
            \STATE $\mathbf{V} \leftarrow \mathrm{findVertices}(\bar{\mathcal{P}}_{-1})$ \label{alg:one_step_backprojection:vrep}
            \STATE $\CROWNA, \CROWNb \leftarrow \mathrm{CROWNSimplex}(p(\cdot; \pi), \mathbf{V}, \mathbf{A}_{\mathcal{X}_T})$ \label{alg:one_step_backprojection:crown}
        \ELSIF{$alg == \text{DRIP-HPoly}$}
            \STATE $\bar{\mathcal{P}}^\text{rect}_{-1} \leftarrow \mathrm{findRectangleBounds}(\bar{\mathcal{P}}_{-1})$
            \STATE $\CROWNA, \CROWNb \leftarrow \mathrm{CROWN}(p(\cdot; \pi), \bar{\mathcal{P}}^\text{rect}_{-1}, \mathbf{A}_{\mathcal{X}_T})$ \label{alg:one_step_backprojection:crown_orig}
        \ENDIF
        \STATE $\bar{\mathcal{P}}_{-1} \leftarrow \left\{\mathbf{x}\ |\ \begin{bmatrix} \CROWNA \\ \mathbf{A}_{\bar{\mathcal{P}}_{-1}} \end{bmatrix} \mathbf{x} \leq \begin{bmatrix} \mathbf{b}_T - \CROWNb \\ \mathbf{b}_{\bar{\mathcal{P}}_{-1}} \end{bmatrix}\right\}$\label{alg:one_step_backprojection:new_hrep2}
    \ENDFOR
 \RETURN $\bar{\mathcal{P}}_{-1}$
 \end{algorithmic}\label{alg:one_step_backprojection}
\end{algorithm}
\begin{algorithm}[t]
 \caption{Multi-Step DRIP}
 \begin{algorithmic}[1]
 \setcounter{ALC@unique}{0}
 \renewcommand{\algorithmicrequire}{\textbf{Input:}}
 \renewcommand{\algorithmicensure}{\textbf{Output:}}
 \REQUIRE target set $\mathcal{X}_T$, policy $\pi$, dynamics $p$, iterations $n_\text{iters}$, time horizon $\tau$
 \ENSURE BP set approximations $\bar{\mathcal{P}}_{-\tau:0}(\mathcal{X}_T)$
    \STATE $\bar{\mathcal{P}}_{0}(\mathcal{X}_T) \leftarrow \mathcal{X}_T$ \label{alg:backprojection:initialize}
    \FOR{$t$ in $\{-1, -2, \ldots, -\tau\}$} \label{alg:backprojection:timestep_for_loop}
        \STATE $\bar{\mathcal{P}}_{t}(\mathcal{X}_T) \leftarrow \text{DRIP}(\bar{\mathcal{P}}_{t+1}(\mathcal{X}_T), \pi, p, n_\text{iters})$ \label{alg:backprojection:one_step_backproj_iter}
    \ENDFOR
 \RETURN $\bar{\mathcal{P}}_{-\tau:0}(\mathcal{X}_T)$ \label{alg:backprojection:return}
 \end{algorithmic}\label{alg:backprojection}
\end{algorithm}

\subsection{Algorithm Overview}\label{sec:approach:algorithm}

To summarize an implementation of the proposed approach, we first describe the method for relaxing $p$ (\cref{alg:crown_simplex}) over a simplex domain, then show how to calculate the BPOA for a single timestep (\cref{alg:one_step_backprojection}), then describe how to compute BPOAs over multiple timesteps (\cref{alg:backprojection}).

\begin{table*}
    \centering
    \captionof{figure}{Comparison of BP set bounds. Given a target set (red), proposed methods lead to much tighter bounds (dashed lines) on true BP sets (solid lines), by combining multiple iterations (bottom row), polytope target sets (middle column), and simplex bounds during relaxation (right column).}
    \label{fig:double_integrator_sets}
    \setlength{\tabcolsep}{0pt}
    \begin{tabular}{cM{58mm}M{58mm}M{58mm}}
        & LP solver & Polytope target sets (Proposed) & Polytope target sets + simplex bounds (Proposed)  \\
        \rotatebox[origin=c]{90}{1 Iteration} & 
        \begin{subfigure}{\linewidth}\centering\includegraphics[width=\linewidth]{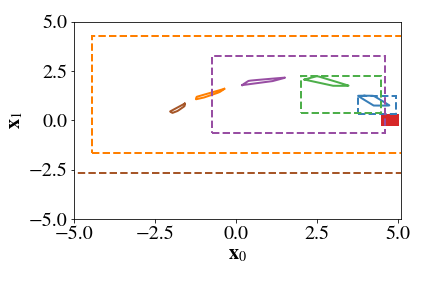}\vspace{-0.2in}\caption{Prior Work: BReachLP~\cite{Rober22_CDC}}\label{fig:double_integrator_sets:lp_1}\end{subfigure}
        & 
        \begin{subfigure}{\linewidth}\centering\includegraphics[width=\linewidth]{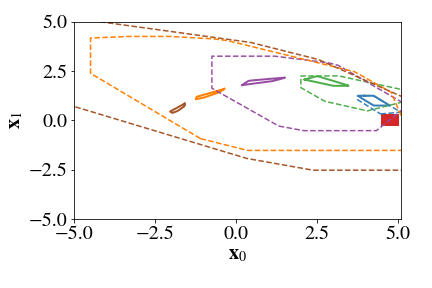}\vspace{-0.2in}\caption{Proposed: DRIP-HPoly}\label{fig:double_integrator_sets:rect_1}\end{subfigure}
        & 
        \begin{subfigure}{\linewidth}\centering\includegraphics[width=\linewidth]{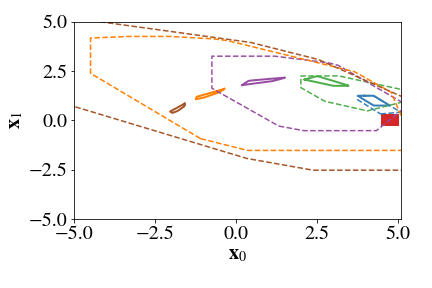}\vspace{-0.2in}\caption{Proposed: DRIP}\label{fig:double_integrator_sets:poly_1}\end{subfigure}
        \\
        \rotatebox[origin=c]{90}{5 Iterations (Proposed)} & 
        \begin{subfigure}{\linewidth}\centering\vspace{-0.2in}\includegraphics[width=\linewidth]{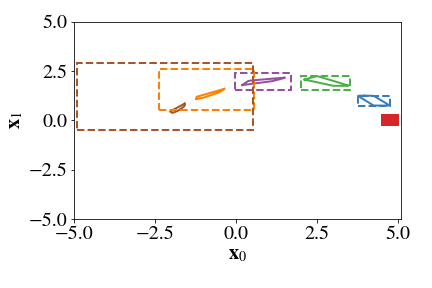}\vspace{-0.2in}\caption{Proposed: BReachLP-Iterate}\label{fig:double_integrator_sets:lp_5}\end{subfigure}
        & 
        \begin{subfigure}{\linewidth}\centering\vspace{-0.2in}\includegraphics[width=\linewidth]{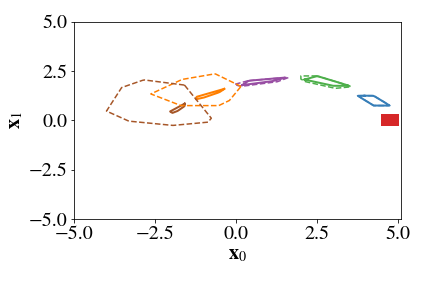}\vspace{-0.2in}\caption{Proposed: DRIP-HPoly}\label{fig:double_integrator_sets:rect_5}\end{subfigure}
        & 
        \begin{subfigure}{\linewidth}\centering\vspace{-0.2in}
        \begin{tikzpicture}
        \node[inner sep=0pt] (plot) at (0,0)
            {\includegraphics[width=\linewidth]{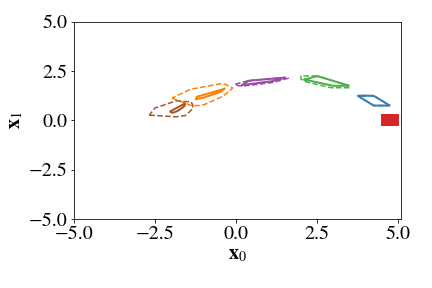}};
        \node[inner sep=0pt,label={[yshift=-0.2cm]Timestep}] (legend) at (0.3,-0.7)
            {\includegraphics[width=0.8\linewidth]{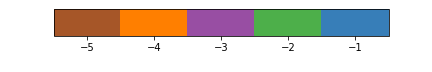}};
        \end{tikzpicture}
        \vspace{-0.3in}\caption{Proposed: DRIP}\label{fig:double_integrator_sets:poly_5}\end{subfigure}
        \\
        \vspace{-0.2in}
    \end{tabular}
    \vspace{-0.3in}
\end{table*}

Given an objective matrix $\mathbf{C}\in \mathds{R}^{h \times n_x}$, \cref{alg:crown_simplex} aims to compute $\mathbf{M}, \mathbf{n}$ such that $\mathbf{C}\mathbf{x}_t \leq \mathbf{M} \mathbf{x}_{t-1} + \mathbf{n}$ $\forall \mathbf{x}_{t-1} \in \texttt{conv\_hull}(\mathbf{V})$.
Add a linear transformation to the front of the closed-loop dynamics to obtain a new function, \mbox{$f_\text{simplex}: \Delta_{n_v} \to \mathcal{X}$} (\cref{alg:crown_simplex:add_layer}).
Then, define the parameters of each relaxation $\alphab_{l,i}, \betab_{l,i}, \alphab_{u,i}, \betab_{u,i}$, from \cref{eq:lin_relaxation_zi}, specifying the standard $n_v$-simplex as the bounds on the function's input (\cref{alg:crown_simplex:fwd_bounds}).
Finally, compute backward bounds from the objective, $\mathbf{C}$, to the current state, $\mathbf{x}_{t}$, stopping before reaching the added linear transformation (\cref{alg:crown_simplex:bwd_bounds}).

\cref{alg:one_step_backprojection} describes the proposed method for computing a BPOA, $\bar{\mathcal{P}}_{-1}$.
First, extract $\mathbf{A}_T, \mathbf{b}_T$ as the H-rep of the target set polytope.
Then, hyper-rectangle $\bar{\mathcal{R}}_{-1}$ is computed by solving $2n_x$ LPs (\cref{alg:one_step_backprojection:backreach}), and the BPOA, $\bar{\mathcal{P}}_{-1}$, is initialized as $\bar{\mathcal{R}}_{-1}$.
To refine that estimate, loop for $n_\text{iters}$ (\cref{alg:one_step_backprojection:loop}).
If using DRIP, at each iteration, compute the vertex representation (V-rep) of the current BPOA (\cref{alg:one_step_backprojection:vrep}, we used~\cite{caron2018python}) and run \cref{alg:crown_simplex} with $\mathbf{C}=\mathbf{A}_T$ on this simplex domain (\cref{alg:one_step_backprojection:crown}).
If using DRIP-HPoly, find hyperrectangle bounds on $\bar{\mathcal{P}}_{-1}$ and run CROWN~\cite{zhang2018efficient} with $\mathbf{C}=\mathbf{A}_T$ on this hyperrectangular domain.
Note that DRIP-HPoly completely avoids the conversion between H-Rep and V-Rep, which may be desirable for higher dimensional systems.
Either way, then set the parameters of the H-rep of the refined BPOA using the CROWN relaxation and target set along with the prior iteration's BPOA.
Update $\bar{\mathcal{P}}_{-1}$ and the loop continues.
After $n_\text{iters}$, $\bar{\mathcal{P}}_{-1}$ is returned.
To calculate BPOAs for a time horizon $\tau$, \cref{alg:backprojection} calls \cref{alg:one_step_backprojection} $\tau$ times, using the next step's BPOA as the target set.

\subsection{Time Complexity Analysis}

First, we analyze the growth of $n_\text{facets}$ from domain refinement iteration.
At $t{=}-1$, the BPOA starts with $n_\text{facets} = 2n_x$ (hyperrectangle, $\bar{\mathcal{R}}_{-1}$).
At each iteration, the BPOA gains at most $2n_x$ facets, meaning $n_\text{facets}{=}2n_x + 2n_\text{iters}n_x$ at the end of the first timestep's domain iteration.
At $t{=}-2$, the BPOA again starts with $n_\text{facets}{=}2n_x$, but at each iteration we add $2n_x + 2n_\text{iters}n_x$ facets, as the target set is the BPOA at $t{=}-1$.
After $T$ steps of $n_\text{iters}$ iterations, the BPOA has at most $O(n_\text{iters}^T n_x)$ facets.

\textbf{DRIP-HPoly:}
Assume solving 1 LP has complexity $O((n+d)^{1.5} n L)$~\cite{vaidya1987algorithm}, with $n$ variables, $d$ constraints, and $L$ encoding bits.
BR bounds require $2 n_x$ LPs, where each has $n_\text{facets} + 2n_u$ constraints and $n_x+n_u$ decision variables.
Assuming $n_\text{facets}{>}2n_u$ and $n_x{>}n_u$, this gives $O((n_x+n_\text{iters}^T n_x)^{1.5} n_x L) = O(n_\text{iters}^{1.5T} n_x^{2.5} L)$ runtime across all $T$ timesteps.
At each refinement iteration, we find rectangular bounds and run CROWN.
Finding rectangular bounds is similar to getting BR bounds except it is done $n_\text{iters}$ times, giving $O(n_\text{iters}^{1.5T+1} n_x^{2.5} L)$.
CROWN time complexity is $O(m^2 n^3)$ for an $m$-layer network with $n$ neurons per layer and $n$ outputs~\cite{zhang2018efficient}; since we have $O(n_\text{iters}^{T-1} n_x)$ rows in the objective, we get $O(m^2 n^3 n_\text{iters}^T n_x)$ (assuming $n{>}n_x$ and $n{>}n_u$).
Relaxing the closed-loop dynamics would be the same ($p$ contains the control NN with a constant number of additional layers).
Thus, to compute BPOAs for $T$ timesteps, DRIP-HPoly's complexity is $O(n_\text{iters}^{1.5T+1} n_x^{2.5} L + m^2 n^3 n_\text{iters}^T n_x)$.

\textbf{DRIP:}
Polytope domain relaxation affects the analysis in 3 places: vertex enumeration (converting from H-rep to V-rep), CROWN relaxation (network width), and LP (number of constraints).
Since each BPOA starts as a hyperrectangle, the polytopes are always bounded.
Assume that vertex enumeration time complexity for a bounded polytope is $O(n^2 d v)$ with $v$ vertices from $n$ hyperplanes in $d$ dimensions~\cite{avis1991pivoting}.
Here, $d=n_x$ and we assume $\Theta(n_\text{facets}^{\floor{\frac{n_x}{2}}})$ (worst-case caused by cyclic polytopes~\cite{toth2017handbook}) vertices, which corresponds to $O(n_\text{facets}^{2+\floor{\frac{n_x}{2}}} n_x)$.
Recall that after $T$ timesteps of $n_\text{iters}$ iterations, the BPOA would have $O(n_\text{iters}^T n_x)$ facets; it could thus have $O(n_\text{iters}^{T\floor{\frac{n_x}{2}}} n_x^{\floor{\frac{n_x}{2}}})$ vertices.
Therefore, the time complexity of vertex enumeration is $O(n_\text{iters}^{T(2+\floor{\frac{n_x}{2}})} n_x^{3+\floor{\frac{n_x}{2}}})$.
Analagously, the CROWN runtime is $O(n_\text{iters} m^2 (n_\text{iters}^{\floor{\frac{n_x}{2}}} n_x^{\floor{\frac{n_x}{2}}})^3 n_\text{iters}^T n_x) = O(m^2 n_\text{iters}^{1+T+3T\floor{\frac{n_x}{2}}} n_x^{3\floor{\frac{n_x}{2}}+1})$, since we add a single layer to the NN with one neuron per polytope vertex.
For the LP runtime, there are $O(n_\text{iters}^{T-1} n_x)$ constraints, leading to $O(n_\text{iters}^{1.5 (T-1)} n_x^{3.5} L)$ time complexity.
The full DRIP algorithm has time complexity of $O(n_\text{iters}^{T(2+\floor{\frac{n_x}{2}})} n_x^{3+\floor{\frac{n_x}{2}}} + m^2 n_\text{iters}^{1+T+3T\floor{\frac{n_x}{2}}} n_x^{3\floor{\frac{n_x}{2}}+1} + n_\text{iters}^{1.5 (T-1)} n_x^{3.5} L)$.

In practice, the runtime may be much better.
For example, $T=1$ for \cref{fig:ground_robot}, which eliminates a source of exponential growth.
Alternatively, rather than allowing the number of facets to grow exponentially, one could simply solve an LP at each iteration/timestep to obtain an constant-size outer bound on the latest BPOA.
While this would certainly reduce the exponential growth in $n_\text{facets}$, our numerical experiments suggest that such a step is not necessary in practice for the systems considered.

\section{Results}

This section demonstrates DRIP on two simulated control systems (double integrator and mobile robot), implemented with the Jax framework~\cite{jax2018github} and \texttt{jax\_verify}~\cite{jaxverify} library.

\subsection{Ablation Study}

\begin{figure}[t]
    \centering
    \begin{tikzpicture}
        \node[inner sep=0pt] (plot) at (0,0)
            {\includegraphics[width=0.6\linewidth]{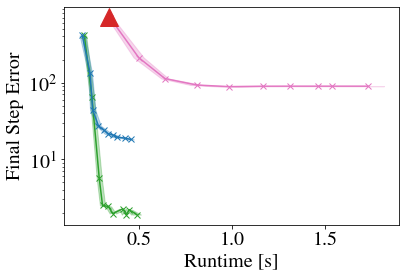}};
        \node[inner sep=0pt] (legend) at (1.0,-0.3)
            {\includegraphics[width=0.3\linewidth]{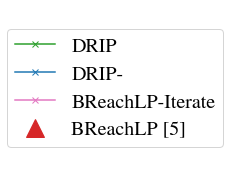}};
        \node at (1.4, -0.15) {\scriptsize{HPoly}};
    \end{tikzpicture}
    \caption{Runtime vs. Error. Compared to prior work (BReachLP), the refinement loop lowers the approximation error with more computation time (BReachLP-Iterate). Incorporating  target set facets reduces the error and runtime (DRIP-HPoly), and incorporating simplex bounds further improves the error with similar runtime (DRIP). Overall, there is a $371\times$ improvement in error for similar runtime (0.3s).}
    \label{fig:runtime_vs_error}
\end{figure}

Using the double integrator system and learned policy from~\cite{everett2021reachability} (2 hidden layers each with 5 neurons), \cref{fig:double_integrator_sets} demonstrates the substantial improvement in reachable set tightness enabled by our proposed method.
For a target set (red) $\mathcal{X}_T = [4.5, 5.0] \times [-0.25, 0.25]$, the true BP sets are shown in solid colors for each timestep, and the BPOAs are shown in the same colors with dashed lines.
For example, \cref{fig:double_integrator_sets:lp_1} implements the prior work~\cite{Rober22_CDC}, in which BPOAs become very conservative after a few timesteps.
By adding the refinement loop (\cref{sec:approach:refinement_loop}), \cref{fig:double_integrator_sets:lp_5} shows substantial improvement with $n_\text{iters} = 5$.

However, increasing $n_\text{iters} > 5$ does not improve the results with the prior LP and hyper-rectangular BPOA formulation.
Instead, the middle column shows the impact of the proposed formulation from \cref{sec:approach:polytope_formulation}, in which the closed-loop dynamics are relaxed to directly provide BPOAs as polytopes.
A key reason behind this improvement is the use of the target set's facets as the objective matrix during the backward pass of the CROWN algorithm.
When increasing $n_\text{iters}$ to 5, we see nearly perfect bounds on the first three BP sets, with tighter yet still somewhat loose bounds on the final two timesteps.

The impact of the polytope domain used in the relaxation (\cref{sec:approach:polytope_input_bounds}) is shown in the rightmost column.
Because the first iteration always uses the hyper-rectangular $\mathcal{R}_{-t}$, we expect to see no difference between \cref{fig:double_integrator_sets:rect_1} and \cref{fig:double_integrator_sets:poly_1}.
However, \cref{fig:double_integrator_sets:poly_5} shows much tighter BPOAs in the last two timesteps when compared to \cref{fig:double_integrator_sets:rect_5}.

Overall, the massive improvement in BPOA tightness between the prior work~\cite{Rober22_CDC} (\cref{fig:double_integrator_sets:lp_1}) and the proposed approach (\cref{fig:double_integrator_sets:poly_5}) would enable a practitioner to certify safety when starting from a larger portion of the state space.

\begin{figure}[t]
    \centering
    \begin{subfigure}{0.45\linewidth}
        \centering
        \includegraphics[width=\linewidth]{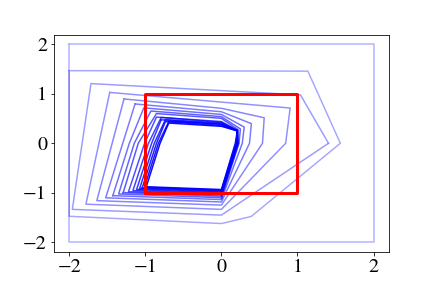}
        \caption{1-step BPOAs (blue) for 15 iterations with obstacle (red)}
        \label{fig:ground_robot:bp}
    \end{subfigure}\hfill
    \begin{subfigure}{0.45\linewidth}
        \centering
        \includegraphics[width=\linewidth]{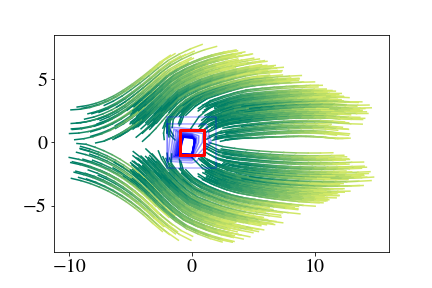}
        \caption{400 rollouts with target set (red) and BPOAs (blue)}
        \label{fig:ground_robot:trajs}
    \end{subfigure}
    \caption{Ground robot policy certification. With $n_\text{iters}=15$, $\bar{\mathcal{P}}_{-1} \subseteq \mathcal{X}_T$, which certifies that the system will never collide with the obstacle (starting anywhere outside the obstacle).}
    \label{fig:ground_robot}
\end{figure}

\subsection{Runtime vs. Error Tradeoff}

\cref{fig:runtime_vs_error} provides quantitative analysis of the tradeoff between tightness and computational runtime for the various methods.
Final step error is the ratio $\frac{A_\text{BPOA} - A_\text{BP}}{A_\text{BP}}$, where $A_{\mathcal{A}}$ denotes the area of set $\mathcal{A}$.
Runtime is plotted as the mean and shaded standard deviation over 20 trials (with the worst one discarded).
For the 3 variants of the proposed algorithms, by increasing $n_\text{iters}$, we observe an improvement in error at the cost of additional runtime.

The red triangle corresponds to \cref{fig:double_integrator_sets:lp_1}~\cite{Rober22_CDC}.
The pink curve corresponds the leftmost column of~\cref{fig:double_integrator_sets}, which adds a refinement loop on the relaxation domain (\cref{sec:approach:refinement_loop}).
The blue curve corresponds to the middle column of~\cref{fig:double_integrator_sets}, which uses polytope BPOAs (\cref{sec:approach:polytope_formulation}).
The green curve corresponds to the rightmost column of~\cref{fig:double_integrator_sets}, which uses polytope relaxation domains (\cref{sec:approach:polytope_input_bounds}). 

A key cause of the computation time reduction between red/pink and green/blue is that the new formulation replaces many of the LPs with closed-form polytope descriptions.
Overall, we observe a $371\times$ improvement in the error.
Moreover, this experiment did not require any set partitioning and was still able to provide very tight BPOAs.

\subsection{Ground Robot Policy Certification}

Next, we use DRIP to certify that a robot will not collide with an obstacle for any initial condition in the state space (outside of the obstacle itself).
We use the 2D ground robot dynamics, policy, and obstacle from~\cite{Rober22_CDC,Rober22_OJCSYS} (2 hidden layers each with 10 neurons).
The target set is partitioned into 4 cells (2 per dimension), which we note is $4\times$ fewer cells than in~\cite{Rober22_CDC,Rober22_OJCSYS}.
For each cell, the 1-step BPOA is calculated for $n_\text{iters}$, and the returned BPOA is the convex hull of the union of vertices for each cell's BPOA.

\cref{fig:ground_robot:bp} shows the target set (red) and BPOAs for $n_\text{iters}=\{0, 1, \ldots, 15\}$ (blue, darker corresponds to larger $n_\text{iters}$).
When $n_\text{iters}=15$, $\bar{\mathcal{P}}_{-1} \subseteq \mathcal{X}_T$.
By Corollary A.2 of~\cite{Rober22_OJCSYS}, this certifies that the system can only reach the obstacle if it starts from within the obstacle; there is no need to compute BPOAs for further timesteps.
\cref{fig:ground_robot:trajs} shows BPOAs with rollouts of the closed-loop dynamics from 400 random initial states, although sampling trajectories is not necessary given the certificate.

\section{Conclusion}

This paper proposed a new backward reachability algorithm, DRIP, for formal safety analysis of data-driven control systems.
DRIP advances the state-of-the-art by introducing ideas to shrink the domain over which the closed-loop dynamics are relaxed and leverage polytope representations of sets at both the input and output of the relaxation.
These innovations are shown to provide 2 orders of magnitude improvement in bound tightness over the prior state-of-the-art with similar runtime.
Future work will investigate tighter simplex bounds~\cite{behl2021overcoming} and convergence properties.

\vspace{-0.1in}
\bibliographystyle{IEEEtran}
\bibliography{refs}

\end{document}